\documentstyle[12pt]{article}
\def\beq{\begin{equation}}
\def\eeq{\end{equation}}
\def\bea{\begin{eqnarray}}
\def\eea{\end{eqnarray}}

\def\ba{\begin{array}}
\def\ea{\end{array}}

\def\,{\"{U}}
\def\6{\.{I}}

\begin{document}
\title{Exact Bound State Solutions of the Schr\"{o}dinger Equation for Noncentral Potential via the Nikiforov-Uvarov Method}

\author{Metin Akta\c{s}\thanks{Corresponding author: E-mail: metin@kku.edu.tr}\\[.5cm]
Department of Physics, Faculty of Arts and Sciences\\
K{\i}r{\i}kkale University, 71450, K{\i}r{\i}kkale, Turkey}

\date{\today}
\maketitle
\normalsize
\begin{abstract}

\noindent Exact bound state solutions of the Schr\"{o}dinger
equation for generalized noncentral potential are examined by means
of the Nikiforov-Uvarov method. The wavefunctions and the
corresponding energy eigenvalues of the system are obtained
analytically. The results examined for the potential are
compatible with those obtained by the other methods.\\

\end{abstract}

\newpage
\section{Introduction}

\bigskip

\noindent Exact analytic studies on the Schr\"{o}dinger equation for
noncentral potentials have been considerable interests in recent
years. Some of the numerous studies concerning in the context are
available in the literature [1-7]. Two different classes of the
potentials are considered in searching both relativistic and
nonrelativistic quantum mechanical systems. The first case is
Coulombic type potential which is used in analyzing the bound states
of an electron within the Hartmann's ring-shaped plus Aharonov-Bohm
(AB) field [8] or a magnetic Dirac monopole [9]. They can be used in
quantum chemistry for describing the ring-shaped organic molecules
such as cyclic polyenes and benzene as well as in nuclear physics
for investigating the interaction between deformed pair of nuclei
and the spin-orbit coupling for the motion of the particle in the
potential fields. The second one is the generalized harmonic
oscillator type potential known also as the ring-shaped oscillator
plus AB-systems [10]. These are very useful for calculating the
vibrational quantum levels and determining the hidden dynamical
symmetries of quantum systems which are described a particle moving
in the potential fields. In this study, we introduce a ring-shaped
type generalized noncentral potential and investigate its solutions
with respect to the Schr\"{o}dinger equation.

\bigskip

\noindent By replacing the Coulomb term of the Hartmann's potential
with the harmonic oscillator one and adding to the inverse square
term (or saying that it is either attractive or repulsive for
negative and positive parameters) respectively. It also includes
contribution of the potential term as $(rcos\theta)^{-2}$. Thus, we
define the potential in spherical coordinates as follows:

\begin{equation}\label{1}
    V_{q}(r,\theta)=\left(A
    r^{2}+\frac{B}{r^{2}}\right)+\frac{q}{r^{2}}\left(\frac{C}{sin^{2}\theta}+\frac{D}{cos^{2}\theta}\right),
\end{equation}

\bigskip

\noindent\vspace{2 mm} where $q$ is the deformation parameter [11].
In equation (1), only $\theta-$dependent part $V(\theta)=$ \vspace{2
mm} $(\frac{C}{sin^{2}\theta}+\frac{D}{cos^{2}\theta})$ is often
called the P\"{o}schl-Teller I type potential with
$C=\frac{V_{0}}{2}\chi(\chi-1)$, \vspace{2 mm}
$D=\frac{V_{0}}{2}\lambda(\lambda-1)$ and
$V_{0}=\frac{\hbar^{2}\alpha^{2}}{m}$ [12]. Various choices of
energy parameters allow us to investigate for some particular cases:

\bigskip

\noindent For $B=0$, the potential (1) turns into the double
ring-shaped oscillator [13]; for $B=D=0$, the ring-shaped oscillator
[11, 13]. Moreover, for only $D=0$, it refers to the nonspherical
ring-shaped oscillator [2]; for only $q=0$ or $C=D=0$, the harmonic
oscillator plus inverse square potentials, respectively. Besides
these, for $B=C=D=0$, the usual spherical harmonic oscillator
potential is possible.

\bigskip

\noindent Noncentral potentials are not usually discussed in quantum
mechanical problems. The main reason is that most of them are not
analytically solvable. In spite of this fact, some classes of
noncentral potentials in three dimensions are solvable so long as
the Schr\"{o}dinger, Klein-Gordon and Dirac equations [14-26] with
these potentials satisfy the separation of variables. These are also
applicable to the scattering and condensed matter processes [27-30].
There have been many studies involving the potentials by using the
well-known techniques, i.e., group theoretical manner [11, 31-33],
supersymmetric formalism [34-41], path integral method [42-48] and
other algebraic approaches [49-58].

\bigskip

\noindent Throughout the study, we have focused on several purposes.
One of them is to solve the separated Schr\"{o}dinger equations for
the potential (1) by using the Nikiforov-Uvarov (NU) method [62]
analytically. Our another interest is to extract some remarkable
results.

\bigskip

\noindent The outline of the paper is as follows: In Sec. II, the
method is introduced briefly. In Sec. III, the wavefunction and the
corresponding energy eigenvalues are obtained analytically. In Sec.
IV, our results are reviewed and discussed.

\section{The Method: Description of the Procedure}
\noindent Let us first introduce the Nikiforov-Uvarov method. Some
quantum mechanical problems can be solved analytically after
applying the separation of variables procedure [7, 49]. Then, a
proper transformation is chosen as $x=x(z)$ for each separated
equation. Therefore, each of them is reduced to a generalized
equation of hypergeometric type as [62]

\bigskip

\begin{equation}\label{3}
    u^{\prime\prime}(z)+\frac{\tilde{\tau}(z)}{\sigma(z)}u^{\prime}(z)
    +\frac{\tilde{\sigma}(z)}{\sigma^{2}(z)}u(z)=0.
\end{equation}

\bigskip

\noindent Here, $\tilde{\tau}(z)$ is a polynomial function with the
highest degree $1$ and $\sigma(z)$ and $\tilde{\sigma}(z)$ are
polynomial functions with the highest degree $2$. It can also be
noted that the energy $E_{n}$ appear as a parameter in these
coefficients. Their solutions, therefore, lead to several classes of
special functions such the classical orthogonal polynomials called
as Hermite, Jacobi, Bessel etc. If we apply the transformation
$u(z)=\phi(z)y(z)$ to original equation (3) and arrange it, we have
a simpler form as

\begin{equation}\label{4}
    \sigma(z)y^{\prime\prime}+\tau(z)y^{\prime}+\lambda y=0.
\end{equation}

\noindent Here, one concludes

\begin{equation}\label{5}
    \frac{\pi(z)}{\sigma(z)}=\frac{d}{dz}[\ln\phi(z)],
\end{equation}

\noindent and

\smallskip

\begin{equation}\label{6}
\tau(z)=\tilde{\tau}(z)+2\pi(z).
\end{equation}

\smallskip

\noindent Moreover, the nontrivial solutions of equation (4) must
satisfy the energy eigenvalue equation

\begin{equation}\label{7}
    \lambda_{n}+n\tau^{\prime}+\frac{1}{2}[n(n-1)\sigma^{\prime\prime}]=0,\qquad
    (n=0, 1, 2\ldots)
\end{equation}

\bigskip

\noindent where~~$\tau^{\prime}$ and $\sigma^{\prime\prime}$ denote
the first and second derivatives of them with respect to $z$. To
determine $\pi(z)$ and $\lambda=\lambda_{n}$ by assuming

\begin{equation}\label{8}
    k=\lambda-\pi^{\prime}(z),
\end{equation}

\bigskip

\noindent and by solving  the quadratic equation for $\pi(z)$, one
gets

\bigskip

\begin{equation}\label{9}
     \pi(z)=\left(\frac{\sigma^{\prime}-\tilde{\tau}}{2}\right)\pm\sqrt{\left(\frac{\sigma^{\prime}
     -\tilde{\tau}}{2}\right)^{2}+k\sigma-\tilde{\sigma}},
\end{equation}

\bigskip

\noindent with the prime factors of $\sigma$ denoted as the
differential at first degree. The polynomial expression $\pi(z)$ has
to be the square of polynomials under the square root sign. The
existence is possible only if its discriminant is zero. Thus, it
takes the form without square root. We also point out that the
determination of $k$ is the essential point in the calculation of
$\pi(z)$. Our problem with a nontrivial solutions corresponds to the
eigenfunctions $y_{n}(z)$ called as the Rodriguez formula

\bigskip

\begin{equation}\label{10}
    y_{n}(z)=\frac{C_{n}}{\rho(z)}\frac{d^{n}}{dz^{n}}[\sigma^{n}(z)\rho(z)],
\end{equation}

\bigskip

\noindent where~~$C_{n}$ and $\rho(z)$ are the normalization
constants and the weight function, respectively. This function must
satisfy the condition

\bigskip

\begin{equation}\label{11}
    \rho(z)\tau(z)=\frac{d}{dz}[\sigma(z)\rho(z)].
\end{equation}

\bigskip

\noindent Therefore, wavefunctions corresponding to discrete
eigenvalues are orthonormal that they comply with

\begin{equation}\label{12}
    \int y_{n}(z)y_{m}(z)\rho(z)dz=\delta_{mn}.
\end{equation}

\bigskip

\noindent If we further specify that the wavefunctions can be
normalized when $m=n$.

\section{Calculations}

\noindent For a particle in any potential field, the wave functions
$\Psi_{n}(\bf{r})$ describe the bound states and the corresponding
energy levels $E_{n}$, then the time-independent Schr\"{o}dinger
wave equation becomes

\begin{equation}\label{2}
    \nabla^{2}\Psi(\textbf{r})+\frac{2m}{\hbar^{2}}\left[E-V(\textbf{r})\right]\Psi(\textbf{r})=0.
\end{equation}

\noindent Let us first consider this equation. By putting the
potential $V(\mathbf{r})$=$V_{q}(r,\theta)$ into the equation in
spherical coordinates, one has

\begin{eqnarray}\label{13}
\frac{1}{r^{2}}\frac{\partial}{\partial
r}\left(r^{2}\frac{\partial\Psi_{E}}{\partial r}\right)
&+&\frac{1}{r^{2}}\left[\frac{1}{\sin\theta}\frac{\partial}{\partial\theta}
\left(\sin\theta~\frac{\partial\Psi_{E}}{\partial\theta}\right)+\frac{1}{\sin^{2}\theta}\frac{\partial^{2}\Psi_{E}}{\partial\varphi^{2}}
\right]\nonumber\\[0.5cm]
&+&\frac{2m}{\hbar^{2}}\left\{E-\left[\left(Ar^{2}+\frac{B}{r^{2}}\right)+\frac{q}{{r}^{2}}\left(\frac{C}{\sin^{2}\theta}
+\frac{D}{{\cos^{2}\theta}}\right) \right]\right\}\Psi_{E}=0 .
\end{eqnarray}

\bigskip

\noindent By taking the total wave function as

\begin{eqnarray}\label{14}
\Psi(\textbf{r})&=&\Psi_{E}(r,\theta,\varphi)\nonumber\\[0.3cm]
&=&\frac{1}{r}R(r)\Theta(\theta)\Phi(\varphi),
\end{eqnarray}

\noindent and substituting it into equation (13), one brings to the
forms

\begin{equation}\label{15}
\frac{d^{2}R}{dr^{2}}+\left[\varepsilon-\left(\bar{A}r^{2}+\frac{\gamma}{r^{2}}\right)\right]R(r)=0,
\end{equation}

\noindent and

\begin{equation}\label{16}
\frac{d^{2}\Theta}{d\theta^{2}}+\cot\theta~\frac{d\Theta}{d\theta}+\left[\Lambda-\left(\frac{\widetilde{C}}{\sin^{2}\theta}
+\frac{\overline{D}}{\cos^{2}\theta}\right)\right]\Theta=0,
\end{equation}

\bigskip

\noindent with the well-known $\varphi$-angle (or azimuthal)
solution

\smallskip

\begin{equation}\label{17}
\Phi_{\bar{m}}(\varphi)=\frac{1}{\sqrt{2\pi}}e^{i\bar{m}\varphi},\qquad
\bar{m}=0,\pm 1,\pm 2\ldots
\end{equation}

\bigskip

\noindent\vspace{1.5mm}Equations (15) and (16) are called radial and
$\Theta$-angle equations, respectively and they are\vspace{1.5mm} in
separated forms. In equations (15) and (16),
$\varepsilon=\varepsilon(A,~B,~\gamma)=\frac{2mE}{\hbar^{2}}$,
$\bar{A}=\frac{2mA}{\hbar^{2}}$,
$\overline{B}=\frac{2mB}{\hbar^{2}}$,
$\overline{C}=\frac{2qmC}{\hbar^{2}}$\vspace{1.5mm}and
$\overline{D}=\frac{2qmD}{\hbar^{^{2}}}$ and also
$\widetilde{C}=\overline{C}+\bar{m}^{^{2}}$,
$\gamma=\overline{B}+\Lambda$ ~ are defined. Here, $\bar{m}^{^{2}}$,
and\vspace{1.5mm}$\Lambda=\Lambda(C,~D,~\bar{m})$ are known as
separation constants. By using the NU-method, we seek for the
solutions of the equations.

\subsection{Energy Eigenvalues}
\subsubsection{Radial equation}

\noindent Let us now consider equation (15) only if $\varepsilon<0$.
When we apply a transformation $z=r^{2}$ to it, then we get the
transformed hypergeometric equation

\smallskip
\begin{equation}\label{18}
R^{\prime\prime}(z)+\frac{1}{2z}R^{\prime}(z)+\frac{1}{4z^{2}}\left[-\bar{A}z^{2}-\varepsilon
z-\gamma\right]R(z)=0.
\end{equation}

\bigskip

\noindent By comparing it with equation (2), one has

\begin{equation}\label{19}
\tilde{\tau}(z)=1,\qquad \sigma(z)=2z,\qquad
\tilde{\sigma}(z)=[-\bar{A}z^{2}-\varepsilon z-\gamma].
\end{equation}

\bigskip

\noindent When we substitute these polynomials into equation (8), we
obtain

\bigskip

\begin{equation}\label{20}
\pi(z)=\frac{1}{2}\pm\frac{1}{2}\sqrt{4\bar{A}z^{2}+4(\varepsilon+2k)z+(1+4\gamma)}.
\end{equation}

\bigskip

\noindent From this equation, $k$ can be determined and we rewrite
it as
\bigskip
\begin{eqnarray}\label{21}
\pi(z)=\frac{1}{2}\pm\frac{1}{2}\left\{\begin{array}{ll}
(2\sqrt{\bar{A}}z-\sqrt{1+4\gamma}),&\mbox{$k=-\frac{\varepsilon}{2}-\frac{1}{2}\sqrt{\bar{A}(1+4\gamma)}$}\\[0.3cm]
(2\sqrt{\bar{A}}z+\sqrt{1+4\gamma}),&\mbox{$k=-\frac{\varepsilon}{2}+\frac{1}{2}\sqrt{\bar{A}(1+4\gamma)}$}.
\end{array}
\right.
\end{eqnarray}

\bigskip

\noindent So the proper are the values of $\pi(z)$ that they must
satisfy the condition $\tau^{\prime}(z)<0$ and we have two cases:\\

$(i)$
$\pi=\pi_{1}(z)=\frac{1}{2}-\frac{1}{2}(2\sqrt{\bar{A}}z-\sqrt{1+4\gamma})$~~
for~~
$k=k_{1}=-\frac{\varepsilon}{2}-\frac{1}{2}\sqrt{\bar{A}(1+4\gamma)}$.\\

\noindent In this time we get

\begin{eqnarray}\label{22}
  \lambda=\lambda_{n} &=&-\frac{\varepsilon}{2}-\frac{1}{2}\sqrt{\bar{A}(1+4\gamma)}-\sqrt{\bar{A}}\nonumber\\[0.2cm]
        &=& 2n\sqrt{\bar{A}}.
\end{eqnarray}

\smallskip

\noindent As a result, it gives us the energy eigenvalue equation

\smallskip

\begin{equation}\label{23}
  E_{n}=\widetilde{A}\left[(2n+1)+\frac{1}{2}\sqrt{1+4\gamma}\right],\qquad
  n=0,~1,~2\ldots.
\end{equation}

\bigskip

\vspace{2mm}\noindent where~~
$\widetilde{A}=\frac{\hbar^{2}}{m}\sqrt{\bar{A}}$.

\bigskip

$(ii)$
$\pi=\pi_{2}=\frac{1}{2}-\frac{1}{2}(2\sqrt{\bar{A}}z+\sqrt{1+4\gamma})$~~
for~~
$k=k_{2}=-\frac{\varepsilon}{2}+\frac{1}{2}\sqrt{\bar{A}(1+4\gamma)}$\\

\noindent By recalculating $\lambda=\lambda_{n}$ and arranging it,
one gets

\bigskip

\begin{equation}\label{24}
E_{n}=\tilde{A}\left[(2n+1)-\frac{1}{2}\sqrt{1+4\gamma}\right],\qquad
  n=0,~1,~2\ldots.
\end{equation}

\bigskip

\noindent We note that these two eigenvalue solutions are obtained
by following the conditions in equation (21). The first one refers
to the solutions of $3-$dimensional harmonic oscillator, by taking
care of the parameters given below of equation (17) as $B=0$ and
$\gamma\equiv\Lambda=\ell(\ell+1)$.

\subsubsection{$\Theta$-Angle~Equation}

\noindent  In order to solve equation (16) by the NU-method, we need
to recast it into a new solvable form. Thus we introduce

\begin{equation}\label{25}
y^{\prime\prime}(x)+f(x)y^{\prime}(x)+g(x)y(x)=0,
\end{equation}

\bigskip

\noindent with~~$y(x)=v(x)p(x)$. Then, it is transformed to

\bigskip

\begin{equation}\label{26}
    v^{\prime\prime}(x)+\left(2\frac{p^{\prime}}{p}+f\right)v^{\prime}(x)+\frac{1}{p}(p^{\prime\prime}+fp^{\prime}+gp)v(x)=0,
\end{equation}

\bigskip

\noindent\vspace{1.3mm}where we have
$p(x)=\exp\left[-\frac{1}{2}\int f(x) dx\right]$. By applying the
above procedure for equation (16) and taking $v\rightarrow \Theta$,
one yields

\smallskip

\begin{equation}\label{27}
    \Theta^{\prime\prime}(\theta)+\left[\Gamma-(\kappa~cosec^{2}\theta+\overline{D}~sec^{2}\theta)\right]\Theta(\theta)=0,
\end{equation}

\bigskip

\noindent\vspace{1.3mm}where~~
$\Gamma=(\emph{L}+1/2)^{2}=(\Lambda+1/4)$, $\emph{L}$ is used
instead of $\ell$, and $\kappa=(\widetilde{C}-1/4)$. By introducing
a new variable as $\sqrt{t}=\sin\theta$, equation (27) now reads

\bigskip

\begin{equation}\label{28}
    \Theta^{\prime\prime}(t)+\frac{\left(\frac{1}{2}-t\right)}{t(1-t)}\Theta^{\prime}(t)+\frac{1}{[t(1-t)]^{2}}
    \left[\widetilde{\Gamma}~t(1-t)-\widetilde{\kappa}(1-t)-\widetilde{D}t\right]\Theta(t)=0,
\end{equation}

\bigskip

\vspace{1.7mm}\noindent Here, we have $\widetilde{\Gamma}=\Gamma/2$,
$\widetilde{\kappa}=\kappa/4$ and $\widetilde{D}=\overline{D}/4$.
Again comparing it with equation (2), one has

\bigskip

\begin{equation}\label{29}
\tilde{\tau}(t)=\left(\frac{1}{2}-t\right),\qquad
\sigma(t)=t(1-t),\qquad
\tilde{\sigma}(t)=\left[-\widetilde{\Gamma}~t^{2}+\zeta
t-\widetilde{\kappa}\right],
\end{equation}

\bigskip

\noindent with~~
$\zeta=\left(\widetilde{\Gamma}+\widetilde{\kappa}-\widetilde{D}\right)$.
By substituting these polynomials into equation (8), one gets

\bigskip

\begin{equation}\label{30}
\pi(t)=\frac{1}{4}(1-2t)\pm\frac{1}{4}\sqrt{a_{1}t^{2}+b_{1}t+c_{1}}.
\end{equation}

\bigskip

\noindent Here,~~$a_{1}$, $b_{1}$ and $c_{1}$ parameters are equal
to $(4+16\widetilde{\Gamma}-16\widetilde{\kappa})$,
$(4-16\zeta+16k)$ and $(1+16\widetilde{\kappa})$, respectively. From
this equation, $k$ is determined and we rewrite it

\begin{eqnarray}\label{31}
&&\hspace{-0.6cm}\pi(t)=\frac{1}{4}(1-2t)\pm\frac{1}{4}\nonumber\\[0.2cm]
&&\hspace{-0.6cm}\left\{\begin{array}{cc}
\left[\sqrt{(1+4\kappa)}+\sqrt{(1+4\overline{D})}\right]t-\sqrt{(1+4\kappa)},
&\mbox{$k=\frac{1}{8}+\frac{1}{4}\left[\Gamma-(\kappa+\overline{D})\right]
-\frac{1}{8}\sqrt{(1+4\kappa)(1+4\overline{D})}$}\\[0.8cm]
\left[\sqrt{(1+4\kappa)}-\sqrt{(1+4\overline{D})}\right]t-\sqrt{(1+4\kappa)},
&\mbox{$k=\frac{1}{8}+\frac{1}{4}\left[\Gamma
-(\kappa+\overline{D})\right]+\frac{1}{8}\sqrt{(1+4\kappa)(1+4\overline{D})}$}.
\end{array}
\right.
\end{eqnarray}

\bigskip

\noindent Proper values of $\pi(t)$ can be chosen so that they must
satisfy $\tau^{\prime}(t)<0$. Two possible cases are valid as
follows:

\bigskip

$(i)$\vspace{2mm} Having
$\pi=\pi_{1}(t)=\frac{1}{4}(1-2t)-\frac{1}{4}\left\{\left[\sqrt{(1+4\kappa)}
+\sqrt{(1+4\overline{D})}~\right]t-\sqrt{(1+4\kappa)}\right\}$~~~for
$k=k_{3}=\frac{1}{8}+\frac{1}{4}\left[\Gamma
-(\kappa+\overline{D})\right]-\frac{1}{8}\sqrt{(1+4\kappa)(1+4\overline{D})}$,
we get

\smallskip

\begin{eqnarray}\label{32}
  \lambda=\lambda_{\bar{n}} &=&\bar{n}^{2}+\bar{n}\left\{1+\frac{1}{2}\left[\sqrt{(1+4\kappa)}
  +\sqrt{(1+4\overline{D})}~\right]\right\}\nonumber\\[0.2cm]
        &=& -\frac{3}{8}+\frac{1}{4}\left[\Gamma
-(\kappa+\overline{D})\right]-\frac{1}{8}\sqrt{(1+4\kappa)(1+4\overline{D})}\nonumber\\[0.2cm]
&&-\frac{1}{4}\left[\sqrt{(1+4\kappa)}+\sqrt{(1+4\overline{D})}\right].
\end{eqnarray}

\smallskip

\noindent By putting the potential parameters cautiously into the
equation and rearranging its right hand sides only, one obtains

\begin{eqnarray}\label{33}
  \Lambda&\equiv&\emph{L}(\emph{L}+1)\nonumber\\[0.2cm]
        &=&\left(s_{1}-\frac{1}{2}\right)\left(s_{1}+\frac{1}{2}\right),
\end{eqnarray}

\bigskip

\noindent with the compact parameter

\bigskip

\begin{displaymath}
\hspace{0.01cm}
s_{1}=\sqrt{(2\bar{n}+1)\left[(2\bar{n}+1)+\left(\sqrt{(1+4\kappa)}+\sqrt{(1+4\overline{D})}\right)\right]
+\frac{1}{2}\left[1+\sqrt{(1+4\kappa)(1+4\overline{D})}\right]+(\kappa+\overline{D})}.
\end{displaymath}

\bigskip

$(ii)$ The other solution is

\vspace{2mm}

\begin{displaymath}
\hspace{-3.5cm}\pi=\pi_{2}(t)=\frac{1}{4}(1-2t)-\frac{1}{4}\left\{\left[\sqrt{(1+4\kappa)}
-\sqrt{(1+4\overline{D})}~\right]t-\sqrt{(1+4\kappa)}\right\}
\end{displaymath}

\noindent for~~~$k=k_{4}=\frac{1}{8}+\frac{1}{4}\left[\Gamma
-(\kappa+\overline{D})\right]+\frac{1}{8}\sqrt{(1+4\kappa)(1+4\overline{D})}$,
then we have

\smallskip

\begin{eqnarray}\label{34}
  \Lambda&\equiv&\emph{L}(\emph{L}+1)\nonumber\\[0.2cm]
        &=&\left(s_{2}-\frac{1}{2}\right)\left(s_{2}+\frac{1}{2}\right),
\end{eqnarray}

\noindent with the compact parameter

\bigskip

\begin{displaymath}
\hspace{0.01cm}s_{2}=\sqrt{(2\bar{n}+1)^{2}-(2\bar{n}+1)\left[\sqrt{(1+4\kappa)}-\sqrt{(1+4\overline{D})}\right]
+\frac{1}{2}\left[1-\sqrt{(1+4\kappa)(1+4\overline{D})}\right]+(\kappa+\overline{D})}.
\end{displaymath}

\bigskip

\noindent Within the framework of NU-method, the procedure which
solutions are considered physically acceptable is important. As the
last two equations satisfy the rule of $\tau^{\prime}(t)<0$ [62],
both are acceptable solutions.

\subsection{Wavefunctions}

\subsubsection{Radial wavefunction}

 \noindent To determine the wavefunctions, we first consider equation (4) and we get

\bigskip

\begin{equation}\label{35}
\phi(z)=z^{\delta/4}~e^{-\frac{1}{2}\sqrt{\bar{A}}z},
\end{equation}

\bigskip

\vspace{1.5mm}\noindent where~~$\delta=(1+\sqrt{1+4\gamma})$ and
$\sqrt{z}=r$. From equation (10) by calculating $\rho(z)$ and
inserting it into equation (9), it stands for the generalized
Laguerre polynomials as

\bigskip

\begin{equation}\label{36}
    y_{n}(z)=\tilde{C}_{n}~\emph{L}_{n}^{^{\mu}}(z),
\end{equation}

\bigskip

\noindent with~~$\mu=(\delta-1)/2$ and the weight function is
$\rho(z)=z^{(\delta-1)}e^{-\sqrt{\bar{A}}z}$. Hence, the radial wave
functions become

\begin{eqnarray}\label{37}
    R(z)&=&\phi(z)y_{n}(z)\nonumber\\[0.2cm]
 &=&\tilde{C}_{n}~z^{\delta/4}~e^{-\frac{1}{2}\sqrt{\bar{A}}z}~\emph{L}_{n}^{^{\mu}}(z),
\end{eqnarray}

\noindent where~~$\tilde{C}_{n}$ are normalization constants. When
we insert the weight function $\rho(z)$ and equation (36) in
equation (11), we obtain

\begin{displaymath}
|\tilde{C}_{n}|=\left(\frac{(-1)^{-n}~(n!)~(\bar{A})^{(\delta-1)/4}}{\sqrt{(n+\delta-1)!}}\right),\qquad
n=0, 1, 2,\ldots
\end{displaymath}

\subsubsection{$\Theta$-Angle Wavefunction}

\noindent From equation (4), one finds

\smallskip

\begin{equation}\label{38}
\phi(t)=[t(1-t)]^{(1+\frac{\Delta}{2})/2},
\end{equation}

\bigskip

\vspace{1.2mm}\noindent with~~
$\Delta=\left[\sqrt{(1+4\kappa)}+\sqrt{(1+4\overline{D})}\right]$.
By obtaining $\rho(t)$ via equation (10) and substituting it into
equation (9), it stands for the Jacobi polynomials as

\bigskip

\begin{equation}\label{39}
    y_{\bar{n}}(t)=\tilde{C}_{\bar{n}}~\emph{P}_{\bar{n}}^{^{(\nu_{1},~\nu_{2})}}(t),
\end{equation}

\bigskip

\noindent\vspace{2mm}where~~$\nu_{1}=\sqrt{(1+4\kappa)}$,
$\nu_{2}=\sqrt{(1+4\overline{D})}$ and $\tilde{C}_{\bar{n}}$ are
normalization constants. Also, $\rho(t)=[t(1-t)]^{\Delta/2}$ is
used. Thus we can write

\smallskip

\begin{eqnarray}\label{40}
    \Theta(t)&=&\phi(t)y_{\bar{n}}(t)\nonumber\\[0.2cm]
    &=&\tilde{C}_{\bar{n}}~[t(1-t)]^{(1+\frac{\Delta}{2})/2}
    ~\emph{P}_{\bar{n}}^{^{(\nu_{1},~\nu_{2})}}(t),
\end{eqnarray}

\bigskip

\noindent with~~$\sqrt{t}=\sin\theta$. By means of the similar
procedure as in radial part, we get\\

\begin{displaymath}
|\tilde{C}_{\bar{n}}|=\frac{(2\bar{n}+\Delta)!}{(\bar{n}
+\frac{\Delta}{2})!}\sqrt{\frac{(2\bar{n}+\Delta+1)}{(\bar{n}!)~(\Delta!)}},\qquad
\bar{n}=0, 1, 2,\ldots
\end{displaymath}

\subsubsection{Total Wavefunctions}

\noindent By following equation (14) and gathering equations (17),
(37), (40) respectively, one can construct the total wave function
in compact form

\smallskip

\begin{equation}\label{41}
\Psi(z,t,\varphi)=\widetilde{C}_{n,\bar{n}}~z^{(\delta-1)/4}
~e^{-\frac{1}{2}\sqrt{\bar{A}}z}~\emph{L}_{n}^{^{\mu}}(z)~~
[t(1-t)]^{(1+\frac{\Delta}{2})/2}~
    \emph{P}_{\bar{n}}^{^{(\nu_{1},~\nu_{2})}}(t)~e^{i\bar{m}\varphi},
\end{equation}

\bigskip

\noindent or it can also be written as

\smallskip

\begin{equation}\label{42}
\Psi(r,\theta,\varphi)=\widetilde{C}_{n,\bar{n}}~r^{(\delta-1)/2}~
e^{-\frac{1}{2}\sqrt{\bar{A}}r^{2}}~\emph{L}_{n}^{^{\mu}}(r^{2})~~
(\sin\theta \cos\theta)^{(1+\frac{\Delta}{2})}~
    \emph{P}_{\bar{n}}^{^{(\nu_{1},~\nu_{2})}}(\sin^{2}\theta)~e^{i\bar{m}\varphi},
\end{equation}

\bigskip
\noindent
with~~$\widetilde{C}_{n,\bar{n}}=\frac{|\tilde{C}_{n}|~|\tilde{C}_{\bar{n}}|}{\sqrt{2\pi}}$.

\newpage
\section{Concluding Remarks}

\bigskip

\noindent In this paper, bound-state solutions of the
Schr\"{o}dinger equation for $(r, \theta)$ dependent generalized
noncentral potential have been investigated by the Nikiforov-Uvarov
method. Both the wavefunctions and the corresponding energy spectra
of the system have an exact and explicit forms. Some remarkable
results are noted. By various choices of potential parameters, our
results can also be reduced to the solutions of some quantum
mechanical systems. Possible cases and results are as follows:

\bigskip

\indent $(i)$ For example, equation (23)  turns into the energy
eigenvalues of the spherical harmonic oscillator as

\begin{displaymath}
E_{n}=\hbar\omega\left(2n_{r}+\ell+\frac{3}{2}\right),
\end{displaymath}

\vspace{1.2mm}\noindent for cases $A=\frac{m\omega^{2}}{2}$,
$B=C=D=0$ and $\gamma=\Lambda=\ell(\ell+1)$, where $2n_{r}+\ell=n$
(i.e., called as the principal quantum number). It is noted that the
energy levels of the system are degenerate except for the ground
state $n=0$. The degeneracy of the system is increased by the larger
values of the radial quantum number $n_{r}$ as well as the orbital
angular momentum quantum number $\ell$.

\bigskip

\indent $(ii)$  The deformation parameter $q$, in equation (1),
takes some values as [11]. If we take $q=0$ or $C=D=0$ in equation
(1), we can thus get the solutions of harmonic oscillator plus
inverse square potential. To run the procedure, we should follow the
equations from (13) to (16) and consider the parameters in Section
III as well as the equation (23), our result becomes

\bigskip

\begin{displaymath}
E_{n}=\widetilde{A}\left[(2n+1)+\sqrt{\left(\ell+\frac{1}{2}\right)^{2}+\frac{2mB}{\hbar^{2}}}~\right].
\end{displaymath}

\bigskip

\indent $(iii)$ For $q=1$, our results of the ring-shaped oscillator
potential for $(B=D=0)$ and that of axially symmetric potential for
$B=0$ are identical to those of the systems in [11, 13],
respectively.

\bigskip

\indent $(iv)$ If all the energy parameters are chosen greater than
zero, the eigenvalue equation is therefore written with equation
(33) or equation (34)

\bigskip

\begin{displaymath}
E_{n}=\widetilde{A}\left[(2n+1)+\sqrt{1+4\Lambda+\frac{2mB}{\hbar^{2}}}~\right].
\end{displaymath}

\bigskip

\noindent This is the most general energy eigenvalue result of the
system.

\bigskip

\indent $(v)$ When $A=0$, in equation (1), the bound state equation
(23) vanishes or collapses to zero. It implies the second type bound
state (called as pseudo-bound state) for nonlocal interactions in
three-body or N-body systems [59-61]. For $A\rightarrow 0$, the
energy spectrum (23) may however approach to a continuum bound state
(BCS) depending on the coupling parameters $B,~C,~D$ and the
azimuthal quantum number $(\bar{m})$. The occurrence of a BCS is
hence only a necessary but not a sufficient condition for such an
unphysical collapse in few body systems.

\bigskip

\vspace {1mm} \indent $(vi)$ Equation (15) at $r=0$; equations (16),
(27) at $\theta=0$ and $\theta=\pi/2$; and equation \vspace {1mm}
(28) at $t=0$ and $t=1$~~have singularities. It implies that their
solutions are analytic for \vspace {1mm} $t \epsilon[0,~1]$, and
either $\theta \epsilon[0,~\pi/2]$ or $\theta \epsilon[\pi/2,~\pi]$
intervals. Also, the wavefunctions (41) and (42) vanish at points
$z=0$, $t=0$ and $t=1$; $r=0$ and $\theta=0,~\pi/2$, respectively.
As\vspace {1mm} a result, radial and angular components
$(r,~\theta)$ and transformation parameters $(z,~t)$~~in
wavefunctions should not take these values.

\bigskip

\indent $(vii)$ By introducing the appropriate transformation, we
get the solvable form of the  $\Theta-$ angle equation (16) which is
transformed to the equation (27) and (28). Then, $\theta-$dependent
part solutions of (27) refer to bound-state solutions of the
P\"{o}schl-Teller potential I by\vspace {1mm} putting
$\Gamma\rightarrow E$, $\kappa\rightarrow
\frac{V_{0}}{2}\chi(\chi-1)$, $\bar{D}\rightarrow
\lambda(\lambda-1)$ and by mapping $\theta\rightarrow x$  in [12].

\newpage
\noindent $\bf{Acknowledgment}$\\ This work is supported by the
Turkish Scientific and Technological Research Council (TUBITAK).

\end{document}